\documentclass[aps,twocolumn,floats,prl,nofootinbib,superscriptaddress,10pt]{revtex4-1}
\usepackage[utf8]{inputenc}
\usepackage{bm}
\usepackage{amsmath}
\usepackage{comment}
\usepackage{color}

\usepackage{natbib, float, graphicx,amsmath,amsfonts,amssymb}
\usepackage[utf8]{inputenc}

\def\VEV#1{{\left\langle #1 \right\rangle}}

\begin{document}
\title{Searching for Decaying and Annihilating Dark Matter with
Line Intensity Mapping}

\author{Cyril Creque-Sarbinowski and Marc Kamionkowski}
\email{creque@jhu.edu}
\email{kamion@jhu.edu}
\affiliation{Department of Physics and Astronomy, Johns Hopkins
     University, 3400 N.\ Charles St., Baltimore, MD 21218, USA}

\begin{abstract}
The purpose of line-intensity mapping (IM), an emerging tool for
extragalactic astronomy and cosmology, is to measure the
integrated emission along the line of sight from spectral lines
emitted from galaxies and the intergalactic medium.  The
observed frequency of the line then provides a distance
determination allowing the three-dimensional distribution of the
emitters to be mapped.  Here we discuss the possibility to use these
measurements to seek radiative decays or annihilations from dark-matter
particles.  The photons from monoenergetic decays will be
correlated with the mass distribution, which can be determined
from galaxy surveys, weak-lensing surveys, or the IM mapping
experiments themselves.  We discuss how to seek this
cross-correlation and then estimate the sensitivity of various
IM experiments in the dark-matter mass-lifetime parameter
space.  We find prospects for improvements of ten orders of
magnitude in sensitivity to decaying/annihilating dark matter
in the frequency bands targeted for IM experiments.
\end{abstract}

\maketitle

Measurements of the cosmic microwave background (CMB)
temperature and polarization angular power spectra agree at the
percent level with the predictions of $\Lambda$CDM, a
six-parameter phenomenological model \cite{1502.01589,
1212.5226}. However, the nature of the cold dark matter required
by the model remains a mystery.  It could be primordial black
holes \cite{Carr:1974nx,1603.00464,1603.05234}, axions
\cite{Turner:1989vc,Raffelt:1990yz,1110.2895,1510.07633,Carosi:2013rla},
sterile neutrinos \cite{0906.2968},
weakly interacting massive particles
\cite{hep-ph/9506380,hep-ph/0002126,hep-ph/0701197}, something
related to baryons \cite{1308.0338}, or
any of a rich array of other possibilities \cite{hep-ph/0404175}.
There is, however, no prevailing frontrunner among this vast
assemblage of ideas, and so any empirical avenue that might
provide some hint to the nature of dark matter should be
pursued.

In some scenarios a feeble, but nonzero, electromagnetic
coupling of the dark-matter particle allows it to decay to a
photon line.  For example, the axion undergoes two-photon decay,
and there are ideas (e.g., axion-mediated dark-photon
mixing \cite{1611.01466,1609.06623,1806.09508,1708.04253}) in which the
standard axion phenomenology may be extended.
Monoenergetic photons may be emitted in
the decay of sterile-neutrino dark matter \cite{0906.2968}, and
there are other ideas (e.g., exciting dark matter
\cite{Finkbeiner:2007kk}) in which two dark-matter states are
connected by emission of a photon of some fixed energy.  
There
are also an array of possibilities for monoenergetic photons to
be produced in dark-matter annihilation.  These possibilities
have fueled an extensive search for cosmic-background photons
from dark-matter decays or annihilations at an array of
frequencies through an array of techniques, among them searches
in the extragalactic background light
\cite{Ressell:1991zv,Kamionkowski:1994je,Overduin:2004sz}.

In this paper, we propose to use line-intensity mapping (IM)
\cite{astro-ph/0608032,0910.3010,1709.09066}, an
emerging technique in observational cosmology, to seek radiative
dark-matter decays in the extragalactic background light.
Intensity-mapping experiments measure the
brightness of a given galactic emission line as a function of
position on the sky and observer frequency (which provides a
proxy for the distance) to infer the three-dimensional
distribution of the emitters.  A considerable intensity-mapping
effort with neutral hydrogen's 21-cm line is already underway
\cite{1406.2288}, and efforts are now afoot to develop analogous
capabilities with CO and CII molecular lines, hydrogen's
H$\alpha$ line (e.g., with SPHEREx \cite{1412.4872},
now in a NASA Phase A MidEx study), and others.  If dark matter decays to
a line, the decay photons will be correlated with the mass
distribution, which can be inferred from galaxy surveys,
weak-gravitational-lensing maps, or from the intensity-mapping
surveys themselves.

Our work follows in spirit
Refs.~\cite{Bershady:1990sw,Grin:2006aw}, who sought a
cross-correlation of an axion decay line with the mass
distribution within a given
galaxy cluster, but substitutes the cosmic mass distribution for
the mass distribution within an individual cluster.  We extend
on Ref.~\cite{Gong:2015hke} which
sought angular infrared-background-light fluctuations from
dark-matter decay, by our inclusion of frequency dependence and
cross-correlation with the three-dimensional mass
distribution.  We take here a theorist's perspective,
initially exploring possibilities limited only by astrophysical
backgrounds and assuming perfect measurements.  Doing so, we
find the potential for improvements over current sensitivities
of up to nine orders of magnitude, in frequency bands targeted
by forthcoming IM efforts.  We then show that the improvements
in sensitivity are still dramatic even after taking into account
the effects of realistic instrumental noise.

To begin, consider a Big Bang relic of mass $m_\chi$ and decay rate $\Gamma$ and to be concrete, suppose that the particle, like
the axion, decays to two photons, each of frequency
$\nu_0= m_\chi c^2/(2h) = 1.21\times10^{14}\,(m_\chi
c^2/\text{eV})$~Hz (where $h$ is Planck's constant).  The
time dependence of the energy density of decaying-dark-matter
particles  is $\rho_\chi(t)
=\rho_{\chi 0} a^{-3} e^{-\Gamma (t-t_0)}$ with the 
scale factor $a(t)$ (normalized to $a(t_0)=1$), given as a
function of time $t$.  Here $\rho_{\chi 0}=\rho_{\chi}(t_0)$, and $t_0$ is
the time today.
If the particle in question is {\it the} dark matter, then 
$\Gamma \lesssim H_0 \simeq 10^{-18}$~sec$^{-1}$ (where $H_0$ is
the Hubble parameter).  It is also conceivable, though, that the
Big Bang produced some other relic particle that decays with a
lifetime $\tau =\Gamma^{-1} \lesssim H_0^{-1}$ and thus
does not make up all of the dark matter today.  IM can be used to seek
such short-lived particles as well, although we do not consider
this possibility here.

We now calculate the isotropic specific intensity of decay
photons today.  The specific luminosity density (energy density
per unit time per unit frequency interval) due to decays is
$\epsilon_\nu(t) = \Gamma \rho_\chi(t) \theta(\nu-\nu_0)$, where
$\theta(\nu)$ is the line profile of the signal. We take
$\theta(\nu)$ to be the Dirac delta function.  The specific
intensity observed at frequency $\nu$ and $z = 0$ is given by
the solution \cite{1405.0489,1309.2295},
\begin{equation}
     I_{\nu} = \frac{1}{4\pi} \int_0^\infty dz
     \frac{c}{H(z)}\frac{\epsilon_{\nu(1+z)}(z)}{(1 + z)^4},
\end{equation}
to the radiative-transfer equation, where
$H(z)$ is the Hubble parameter at redshift $z$,
and we have used $z$ as a proxy for $t$.  For dark-matter-decay
photons, this expression evaluates to
\begin{equation}
     I_{\nu} = \left. \frac{c}{4 \pi \nu_0}
     \frac{\Gamma}{H_0}\frac{\rho_\chi(z)}{ E(z) (1+z)^4} \right|_{z=\frac{\nu_0}{\nu}-1},
\end{equation}
where $H(z) \equiv H_0 E(z)$, and for $\Lambda$CDM, $E(z) =
\left[ \Omega_m (1+z)^3 + (1-\Omega_m) \right]^{1/2}$, with
$\Omega_m$ the matter-density parameter today.  Using
$\rho_{\chi 0}
= f\Omega_c \rho_c$, where $\rho_c$ is the critical energy
density of the Universe, $\Omega_c$ the fraction of critical
density in dark matter, and $f$ the fraction of dark matter
today in decaying particles, the specific intensity evaluates,
using Planck 2015 parameters, to 
\begin{equation}
     \nu I_{\nu} = 4.8\times 10^{-3}\, {\rm W}\, {\rm m}^{-2}\,{\rm sr}^{-1}
      \left. \frac{\nu}{\nu_0}\frac{\Gamma}{H_0}\frac{f
     e^{-\Gamma(t-t_0)}}{ E(z)\, (1+z)}\right|_{z = \frac{\nu_0}{\nu} - 1},
\label{eqn:numbers}
\end{equation}
where $t$ is the cosmic time at redshift $z$.
For values $\Gamma\lesssim H_0$ considered here, the exponent
can be taken to be $\exp[-\Gamma(t-t_0)]\simeq 1$.  The intensity
is nonzero only for frequencies $\nu\leq\nu_0$.

The specific fluence (number flux of photons, over all
directions, per unit frequency interval) is $F_{\nu} = 4\pi
I_{\nu}/(h \nu)$, and the total fluence (over all photon
directions) is $F = \int\, F_{\nu}\, d\nu$.
For example, for $f = 1$, $m_\chi c^2= 1$~eV, and $\Gamma=H_0$, the
total fluence evaluates to $5.8 \times
10^{17}\ $m$^{-2}$~sec$^{-1}$.

Photons from dark-matter decay must be
distinguished from a huge background of photons in the
extragalactic background.  We propose here to use the
cross-correlation of these decay photons (which trace the
dark-matter distribution) with some tracer of the large-scale
mass distribution to distinguish decay photons from those in the
extragalactic background.

We now calculate the smallest number of decay photons that
need to be detected to establish their correlation with large-scale
structure.  To do so, we assume that we have sampled the
distribution of mass over a cosmological volume $V$ with a
matter tracer population (e.g., galaxies) that has a mean number
density $\bar n_g$ and bias $b$.  The fractional mass-density
perturbation is $\delta(\vec x)$ with Fourier transform $\tilde
\delta(\vec k)=\int d^3x \,
\delta(\vec x) e^{i \vec k\cdot \vec x}$, so that the matter power
spectrum $P(k)$ is then obtained from $\VEV{\tilde\delta(\vec k)
\delta^*(\vec k')} = (2\pi)^3 \delta_D(\vec k-\vec k') P(k)$,
with $\delta_D(\vec{k})$ the $3$-dimensional Dirac delta
function.  The power spectrum for the tracer population is
$b^2P(k)+ \bar n_g^{-1}$, where we have added a Poisson
contribution due to the finite number density of the tracer
population \cite{1309.2295}.

The same matter-density field is also sampled in the IM
experiment by the photons observed from dark-matter decay.  We
surmise that there are $N_\chi =\xi N_\gamma$ such photons that
appear in the experiment, in addition to $N_b = (1-\xi)N_\gamma$
photons from the extragalactic background light (EBL), the
integrated light from galaxies.  Here $N_\gamma$ is the total
number of observed photons, and the EBL has a frequency spectrum
described in
Refs.~\cite{Ressell:1989rz,Overduin:2004sz,Hill:2018trh}. The
galaxies that give rise to this
EBL are a biased tracer of the mass distribution and so will
also be clustered.  However, the
light from a given galaxy is broadly distributed over
frequencies and so the EBL-photon distribution in the IM
angular-frequency space is smoothed.  We therefore suppose
that the background photons are uniformly distributed throughout
the volume.  We also add to the EBL, for measurements at
frequencies $\nu \lesssim 100$~GHz, Galactic synchrotron
radiation, which we model roughly in terms of a brightness
temperature $T_B = 1000\, {\rm K}\, (\nu/100\,{\rm
MHz})^{-2.5}$.

The fractional luminosity density perturbation in the {\it
observed} photon population is then ${\delta}_\gamma =
N_\chi/(N_\chi+N_b) \delta  =\xi \tilde{\delta}$.
It follows that the cross-correlation between the observed photons and
tracers is 
\begin{equation}
     \VEV{\tilde{\delta}_g(\vec k) \tilde{\delta}_\gamma^{*}(\vec k')} =
     (2\pi)^3 \delta_D(\vec k-\vec k')\, \xi b P(k);
\end{equation}
i.e., the photon-tracer cross-correlation has a power spectrum
$P_{g \gamma}(k) = b \xi P(k)$.

To seek this cross-correlation, we take the Fourier
amplitudes $\tilde \delta_g(\vec k)$ (from the galaxy survey) and
$\tilde\delta_\gamma(\vec k)$ (from the IM experiment, for some
nominal decay frequency $\nu_0$) for each wavevector $\vec k$.
Their product then provides an estimator for $P_{g\gamma}(k)$
with variance,
\begin{equation}
    \left[P_{g\gamma}(k) \right]^2 + \left[P_{\gamma\gamma}(k)+\bar
    n_\gamma^{-1} \right] \left[P_{gg}(k)+ \bar n_g^{-1}\right].
\end{equation}
Here, $\bar n_\gamma=N_\gamma/V$ is the number of photons
collected divided by the volume surveyed, and
$P_{\gamma\gamma}(k) = \xi^2 P(k)$.
If we write $P(k)=A k^{n_s}\left[T(k)\right]^2$, in terms of the
scalar spectral index $n_s\simeq 0.96$, the $\Lambda$CDM
transfer function $T(k)$, and amplitude $A$, then the estimator
from this Fourier mode for $P_{g\gamma}(k)$ provides an
estimator for the product $\xi b$, and thus (if $b$ is known)
for $\xi$.  The minimum-variance estimator for $\xi$ is then
obtained by adding the estimators from each $\vec k$ mode with
inverse-variance weighting.

\begin{figure}[]
\includegraphics[width=0.43\textwidth]{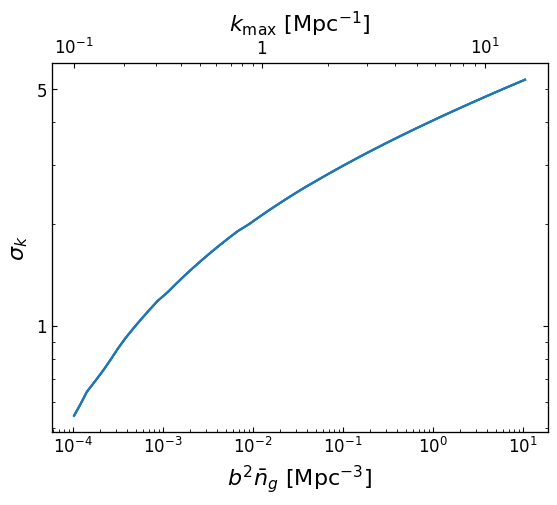}
\caption{The dependence of the mass-density root-variance
     $\sigma_k$ on the tracer-population spatial density $\bar
     n_g$ and its bias $b$.} 
\label{fig:kmax}
\end{figure}

Since the signal for each Fourier mode is $\xi b P(k)$, the squared
signal-to-noise with which the cross-correlation can be measured
with the minimum-variance estimator is
\begin{equation}
     \left( \frac{S}{N} \right)^2 =\sum_{\vec k} \frac{
     \left[\xi b P(k) \right]^2 /2}{\left[\xi b P(k) \right]^2 +
     \left( \xi^2 P(k)+ \bar n_\gamma^{-1} \right) \left(b^2 P(k) +\bar
     n_g^{-1} \right) }.
\end{equation}
To distinguish a detection from the null hypothesis $\xi=0$, we
evaluate the noise under this null hypothesis and
then estimate the sum by approximating $b^2P(k)/\left[b^2
P(k)+\bar n_g^{-1} \right] = 1$ for $b^2 P(k) > \bar
n_g^{-1}$ and $b^2P(k)/\left[b^2
P(k)+\bar n_g^{-1} \right] =0$ for $b^2 P(k) < \bar n_g^{-1}$.
Doing so, we find that the cross-correlation between decay
photons and the tracer population can be measured with a signal
to noise
\begin{align}\label{eqn:S/N}
     (S/N) = \xi\sqrt{N_b
     \sigma_k^2/2},
\end{align}
where $\sigma_k^2 \simeq(2\pi^2)^{-1}
\int^{k_{\rm max}} k^2 P(k)\, dk$ is the variance of the mass
distribution on a distance scale $k_{\rm max}^{-1}$ determined
from $P(k_{\rm max}) = (b^2 \bar n_g)^{-1}$.  Fig.~\ref{fig:kmax}
shows the dependence of $\sigma_k$ on the galaxy number
density $\bar n_g$ and its bias $b$.

Since the scheme suggested here involves {\it cross}-correlation
of a putative decay line with a tracer of the matter
distribution, the measurement is limited {\it not} by the number
of Fourier modes of the density field that can be well sampled,
but only by the fidelity with which the density field can be
sampled and by the number of photons observed.  Moreover, the
factor $\sigma_k^2$ in any measurement is
really the variance of the density field, as determined by the
galaxy survey, in the volume surveyed.

The condition for a $\gtrsim2\sigma$ detection of a decay-line
signal is
\begin{equation}
     \xi \gtrsim \xi_{\rm min} \simeq 2
(N_b \sigma_k^2/2)^{-1/2}.
\label{eqn:xi}
\end{equation}
Since $I_\nu$ from
dark-matter decay and the EBL intensity $I_\nu^{\text{CB}}$ both
vary smoothly with frequency, we approximate $\xi =
N_\chi/(N_\chi+N_b) \simeq N_\chi/N_b$ as $\xi \simeq
I_{\nu}/I_{\nu}^{\text{CB}}$.  We then require, for
detection of a signal,
$I_{\nu_0} \gtrsim \xi_{\rm min} I_{\nu_0}^\text{CB}$, with $I_\nu$
taken from Eq.~(\ref{eqn:numbers}), and evaluated at the
frequency $\nu_0$ at which (for $\Gamma \lesssim H_0$) it peaks.
A cross-correlation can then be detected if the decay lifetime
is 
\begin{align}
     \tau &\equiv \Gamma^{-1} \lesssim
     \frac{ f H_0^{-1} c}{8\sqrt{2}\pi}\sqrt{N_b
     \sigma_k^2}\left[\frac{\Omega_c \rho_c}{\nu_0
     I_{\nu_0}^\text{CB}}\right] \nonumber \\
     &\simeq 7.5\times10^{32} \frac{ f\sigma_k
     (N_b/10^{20})^{1/2}}{(\nu_0
     I_{\nu_0}^\text{CB}/10^{-8}\,\text{W}~\text{m}^{-2}~\text{sr}^{-1})}
     \, \text{sec}.
\label{eqn:lifetimebound}
\end{align}
Here, $N_b$ is the number of EBL photons from the fraction
$f_{\rm sky}$ of the sky observed with frequencies $\nu_1 <\nu <
\nu_2$ that cross a detector area $A$ in time $T$; i.e., 
\begin{align}
     N_b= 4\pi f_\text{sky}AT\int_{\nu_1}^{\nu_2}d\nu\,
     I_\nu^\text{CB}/(h\nu).
\end{align}

We now work out some order-of-magnitude estimates.  Consider
a galaxy survey that samples the volume to redshift $z \sim 1$,
over $f_{\rm sky}$ of the sky.  If the galaxy density is
$\gtrsim 0.01$~Mpc$^{-3}$, then we will have, $k_{\rm max} \gtrsim
0.1$~Mpc$^{-1}$, for which $\sigma_k \gtrsim 1$ (and
we take $b\sim1$).  We consider an intensity-mapping experiment with a
frequency range broad enough to detect decay photons over the
redshift range $0<z<1$, and at frequencies $\nu\sim 10^{14}$ Hz
(roughly optical, probed by SPHEREx, and corresponding to an
dark-matter mass $m_\chi\sim$~eV).  From Fig.~9 in
Ref.~\cite{Hill:2018trh} we ballpark (conservatively) the
EBL intensity as $\nu I_\nu^{\text{CB}} \sim 10^{-8}$~W~m$^{-2}$~sr$^{-1}$
and fluence (over all photon directions) as
$10^{12}$~m$^{-2}$~s$^{-1}$.  We imagine a detector of
area $A_{\rm m}$~m$^{2}$ and observation time $T_{\rm yr}$~yr.   From
Eq.~(\ref{eqn:S/N}), we then estimate the smallest detectable 
$\xi$ to be $\xi_{\rm min}\sim 5\times 10^{-10} (f_{\rm sky} A_{\rm m} T_{\rm
yr})^{-1/2}$ (at $2\sigma$).  Comparing the EBL fluence with our
prior result, $\sim 6\times 10^{17}$~m$^{-2}$~s$^{-1}$, for the
fluence from dark-matter decay with $m_\chi c^2\sim$eV, $f=1$, and
$\Gamma=H_0$, we infer the largest detectable lifetime for
such a measurement to be
\begin{equation}
    \tau_{\rm max} \simeq 10^{33}\, f( f_{\rm sky} A_{\rm m}
    T_{\rm yr})^{1/2} {\rm sec}.
\label{eqn:tauresult}    
\end{equation}
Note that this is $\sim \xi_\text{min}^{-1}\sim 10^9$ times better than the
limit inferred by simply requiring the decay-line intensity to
be smaller than the observed EBL intensity at $\nu_0$ and $\sim 10^8$
times better than the stronger current bounds from null searches
for decay lines from galaxy clusters \cite{Grin:2006aw}.

As Eq.~(\ref{eqn:tauresult}) indicates the sensitivity scales
with the square root of the area on the sky, the area of the
detector, and the duration of the experiment---i.e., with the
square root of the the total number of background photons---as
expected.  There is also a dependence on $\sigma_k$, although
this is weak.  As Fig.~\ref{fig:kmax} shows,
the dependence of $\sigma_k$ with $k_{\rm max}$
is roughly linear for values $k_{\rm max}\sim 0.1$~Mpc$^{-1}$,
and $\tau_{\rm max}$ depends linearly on $\sigma_k$: a greater
density contrast makes a cross-correlation more easily detectable.

The linear-theory calculation
here may be seen as conservative, given that nonlinear evolution
enhances the power spectrum on smaller scales.  On the other
hand, as one goes to smaller scales, the fidelity with which the
galaxy distribution traces the mass distribution decreases.
The numerical estimate above, which used a value
$\sigma_k\sim1$, makes the conservative assumption
that only information from the linear regime is used.

To make the next step in connection to realistic experiments, we now
consider the spatial resolution of the experiment and, moreover,
the anisotropy in the spatial resolution of the cosmic volume
probed. This anisotropy arises because the spatial resolution of the
experiment in the radial direction is fixed by the frequency
resolution; this most generally differs from the spatial
resolution in directions transverse to the line of sight, which
are fixed by the angular resolution. To account for these
effects, we define wavenumbers $k_{\text{max}, ||}$ and
$k_{\text{max}, \perp}$ in the line-of-sight and transverse
directions, respectively. These are fixed by the frequency and
angular resolutions of the experiment. If these wavenumbers are
larger than $k_\text{max}$ from the finite galaxy number density
(discussed above), then the effective
resolution is fixed by finite galaxy number density. If they are
smaller, then the resolution is fixed by the survey
resolutions. That is, we replace the $k_\text{max}$ we derived above by
\begin{align}
\nonumber \tilde{k}_{\max,||} &\equiv \min\left[k_\text{max}, k_{\text{max},||}\sqrt{1 - (k_\perp/k_{\text{max},\perp})^2}\right],\\
\tilde{k}_{\max, \perp} &\equiv \min\left[k_\text{max}, k_{\text{max},\perp}\right],
\end{align}
and then model the Fourier-space volume as an ellipse with the 
corresponding principal semi-axes.  Explicitly, this is
\begin{align}
\nonumber \sigma_k^2 &=
     \frac{1}{2\pi^2}\int^{\tilde{k}_{\text{max},\perp}}\int^{\tilde{k}_{\text{max},||}}
     k_\perp P\left(\sqrt{k_{||}^2 + k_{\perp}^2}\right)
     dk_\perp
     dk_{||}.\\ 
\end{align}
The maximum parallel and
perpendicular wavenumbers are most generally redshift-dependent,
and if so, then we need an integration in Eq.~(\ref{eqn:S/N})
also over cocentric ellipsoidal shells with thickness given by the spectral
resolution and parametrized by redshift.  However, given the
rough nature of our calculation, we approximate the wavenumbers
as constant in redshift as
\begin{align}
     \nonumber k_{\text{max}, ||} &= RH(z)/[c(1 + z)]\approx RH_0/c,\\
     k_{\text{max}, \perp} &= [r_c(z)
     \theta_\text{res}]^{-1}\approx [r_c(0)\theta_\text{res}]^{-1},
\end{align} 
where $r_c(z)$ is the comoving size of the Universe at redshift
$z$, $R$ the spectral resolution, and $\theta_\text{res}$ the
pixel resolution of the experiment.

The last step is to estimate the effects of instrumental noise
on the measurement.  We do so by supposing that instrument noise
distributes $N_{\text{n}}$ photons, in addition to the $N_b$ EBL photons,
uniformly in the survey volume.  The fraction $\xi$ of decay
photons from the sky now gets replaced by a fraction $\xi_{\rm
obs}= N_\chi/(N_\chi+N_b+N_{\text{n}}) = \xi (N_\chi+N_b)/(N_\chi +N_b+N_{\text{n}})$ of the
observed (decay plus EBL plus instrumental-noise) photons that
come from decays.  Following the same reasoning as above, the
smallest detectable $\xi_{\rm obs}$ is $\xi_{\rm obs}^{\rm min}
= 2\sqrt{2}\sqrt{N_b+N_{\text{n}}}/(N_b \sigma_k)$.  The expression,
Eq.~(\ref{eqn:xi}), for the smallest detectable $\xi$, then
becomes,
\begin{equation}
     \xi_{\rm min} \simeq 2 (N_b \sigma_k^2/2)^{-1/2}\sqrt{1+
     (N_{\text{n}}/N_b)},
\label{eqn:ximin}
\end{equation}
after taking into account the $N_{\text{n}}$ additional noise photons.
We can estimate the ratio $(N_{\text{n}}/N_b) \simeq (I_\nu^{\rm
n}/I_\nu^{\text{CB}})$ in terms of an instrument-noise intensity
$I_\nu$ which can be parameterized in terms of the Planck
function $B_\nu(T_{\rm eff})$ at a given effective system
temperature $T_{\rm eff}$.  As Eq.~(\ref{eqn:ximin}) indicates,
if $I_{\nu_0}^{\rm n} \lesssim I_{\nu_0}^{\text{CB}}$, then instrument
noise does not degrade the sensitivity.  
if $I_{\nu_0}^{\rm n} \gtrsim I_{\nu_0}^{\text{CB}}$, then instrument
noise reduces the smallest detectable $\tau$, given in
Eq.~(\ref{eqn:lifetimebound}), by a factor
$(I_{\nu_0}^{\text{CB}}/I_{\nu_0}^{\rm n})^{1/2}$.

We now use Eq.~(\ref{eqn:lifetimebound}),
the cosmic background photon distribution in
Ref.~\cite{Hill:2018trh}, and the experimental parameters shown in 
Table~\protect\ref{table:exp} to forecast lifetime sensitivities for an array
of intensity-mapping experiments that are being pursued or under
consideration.
The array of experiments is chosen to illustrate the different
frequencies (and the principle target astrophysical emission
lines) being targeted by current intensity-mapping efforts.  The
representative efforts are CHIME \cite{1406.2288}, CCAT-prime
\cite{CCAT}; COMAP \cite{1503.08833}; STARFIRE
\cite{starfire}; and SPHEREx \cite{1412.4872}.  There are,
however, a number of
other projects, and several other lines, in other frequency
windows, that may be targets for IM efforts---see
Ref.~\cite{1709.09066} for a more comprehensive list.
We also caution that the planning for some experiments
is not yet complete, and so the detailed parameters may change.

\begin{figure}[]
\includegraphics[width = 0.43\textwidth]{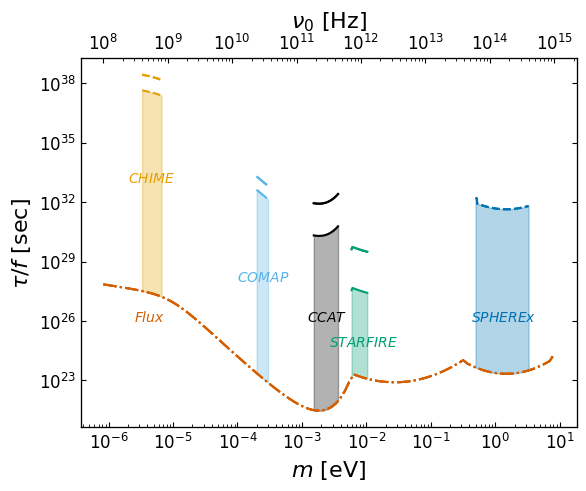}
\caption{The largest lifetime for which a decay signal can be
     detected for the experiments shown.
     The line labeled ``Flux'' is the largest lifetime
     consistent with the requirement that the intensity from
     particle decays does not exceed the extragalactic
     background light (EBL)
     intensity.  The shaded regions are those not yet ruled out
     by current EBL measurements that will be accessible, given
     our estimates of the instrumental noise, with each
     experiment.  The curves above each experiment (except for
     SPHEREx, which is limited by the EBL, not instrument noise)
     show the best sensitivity achievable with an experiment
     with similar specifications, but no instrumental noise, the
     sensitivity being limited in this case by the
     EBL.  The experimental parameters assumed for each
     experiment are listed in Table~\protect\ref{table:exp}.}
     \label{fig:experiments}
\end{figure}

The results are shown in Fig.~\ref{fig:experiments}.  The
experimental parameters we use for these estimates are for the
survey exposure, frequency ranges that are probed, the sky
coverage, and the angular/frequency resolutions.  We
include the dependence of the redshift range probed
on the dark-matter mass and observed frequency range.   We assume that
that $\sigma_k=2.3$ for the tracer population (for $k_{\rm
max}=1.2$~Mpc$^{-1}$, but take into
account the (possibly direction-dependent) modification of
$k_{\rm max}$ determined by the angular and frequency resolution
of the experiment.  For each experiment, we show two curves, one
that includes the estimated effects of instrumental noise, and
another, more optimistic, curve that indicates the limit, from
the EBL, for an experiment with similar exposure and sky and
frequency coverage, but with no instrumental noise.
We also plot in Fig.~\ref{fig:experiments} a current
conservative lower bound to $\tau$ inferred simply by demanding
that the decay-line intensity be less than the observed EBL
intensity at $\nu_0$.

\begin{table*}[htbp]
\begin{tabular}{|c|c|c|c|c|c|c|c|}
\hline
     Experiment & target & $[\nu_1,\nu_2]$ (GHz) & $A$ (m$^2$)&
     $f_\text{sky}$ & $R$ & $\theta_\text{res}$ & $T_{\rm eff}$ K\\
\hline
CCAT & [CII] (high $z$) & $[185, 440]$ & $28$ & $3.9\times 10^{-4}$ & $300$ & $1'$
& 148 \\
CHIME &  21-cm & $[400, 800]\times 10^{-3}$ & $8000$ & $0.75$ &
$\nu/(0.39\ \text{MHz})$ & $20'$ &  100\\
COMAP & CO & $[26, 34]$ &$85$ & $2.4\times 10^{-4}$ & $800$ &
$4'$ & 44 \\
STARFIRE & [CII] (low $z$) & $[714, 1250]$ & $4.9$ & $2.4\times 10^{-5}$ & $250$ &
$1^\circ$ & 11 \\
SPHEREx & H$\alpha$ & $[60, 400]\times 10^3$ & $0.031$ & $1$ & $41.4, 135^*$
& $6.2''$ &  $0^{\dagger}$\\
\hline
\end{tabular}
\caption{The experiment parameters used in
     Fig.~\ref{fig:experiments}.  The survey duration in each
     case was chosen to be one year.  ${}^*$SPHEREx has two
     spectral resolutions, one for the low-frequency channels
     and another at high frequencies.  ${}^\dagger$The noise in
     SPHEREx is limited by zodiacal light and turns out, for our
     purposes, to be negligible compared with that contributed
     by the EBL.}
\label{table:exp}
\end{table*}

As Fig.~\ref{fig:experiments} indicates, IM experiments have the
potential to dramatically increase our ability to seek
radiatively-decaying dark matter in several mass windows, even
with current-generation experiments.  Our discussion above
supposed that the cosmic mass distribution, against which the
decay-line signal is to be cross-correlated, is obtained from a
galaxy survey of number density $\bar n_g$.  However, the mass
distribution is likely to be obtained, over the relevant cosmic
volumes, by the intensity-mapping experiment itself.  If so,
then the relevant value of $\sigma_k$ is that at which the
contribution $P_{\text{n}}(k)$ of the measurement noise (which
may come from shot noise in the emitting sources and/or
instrumental noise) to the power
spectrum overtakes the tracer power spectrum $b^2 P(k)$; i.e.,
$P_{\text{n}}(k_{\rm max})=b^2 P(k_{\rm max})$.  If the decay-line signal
is sought by cross-correlation with a given galactic/IGM
emission line, then our proposed search algorithm resembles
that, proposed in Ref.~\cite{Breysse:2016opl}, to seek a faint
${}^{13}$CO line through cross-correlation with the brighter
${}^{12}$CO line.

We neglected residual correlations in the EBL photons due to
clustering of their sources, since the smoothing of the
galaxy distribution along the line of sight, due to the broad
galaxy spectrum, will suppress these.  We believe that these
residual correlations will be most generally be small, since
the dynamic range of the frequency coverage for most IM mapping
experiments is not too much greater than the width of the galaxy
frequency spectrum.  There may also be spectrum-mapping
techniques, where emission from different frequencies of a
characteristic galaxy spectrum are correlated, that can be used
to mitigate the contamination from EBL clustering.  The
precise sensitivity is likely to be frequency dependent and
weakened considerably if the decay-line frequencies coincides
with that of a strong galactic emission line.

We have chosen, for the purposes of illustration, a scenario
where the dark-matter particle, like the axion, undergoes
two-photon decay.  For axionic dark matter, there is a
constraint $\tau\gtrsim 1.9\times 10^{26} \, (m_a/{\rm
eV})^{-3}$~sec, from horizontal-branch stars, over the mass
ranges considered here.  While this is stronger than the
sensitivities forecast here for $m_a \lesssim 0.1$~eV, it is
considerably weaker than the sensitivity with SPHEREx we
anticipate for $m_a\sim$~eV.  This constraint will not apply
more generally to other particles that undergo radiative decay.
Moreover, while we have discussed line photons from dark-matter
decays, line photons from dark-matter annihilation will be similarly
correlated with the mass (although perhaps with some bias) and
will thus also show up in the search we describe.

To close, we have suggested that lines from dark-matter decay
and annihilation can
be sought by cross-correlating a line signature in
intensity-mapping experiments with some tracer of the mass
distribution.  We have sketched how such a cross-correlation
can be performed and presented simple forecasts for the
sensitivities that can, in principle, be achieved.  Since
intensity mapping is only now getting underway, with long-term
experimental capabilities yet to be specified in detail,
our estimates are for hypothetical experiments limited only by
the unavoidable noise presented by extragalactic background
light.  The potentially extraordinary improvements
to the sensitivity to dark-matter-decay lines we have shown
motivates more careful feasibility studies and adds to the
already strong scientific motivation, from more traditional
astrophysics and cosmology \cite{1412.4872,1709.09066}, to
pursue intensity mapping.

\smallskip
We thank Patrick Breysse and Ely Kovetz for useful comments on a
previous draft.  CS acknowledges the support of the Bill and
Melinda Gates Foundation, Anna Salzberg, and Albert Ratner.
This work was supported by NASA Grant No.\ NNX17AK38G, NSF Grant
No.\ 0244990, and the Simons Foundation.

\end{document}